\documentclass[prl,aps,twocolumn]{revtex4}

\begin{document}

\draft
\title{		Exact Scale Invariance of Composite-Field Coupling Constants}
\author{			Keiichi Akama$^1$ and Takashi Hattori$^2$}
\address{	$^1$Department of Physics, Saitama Medical College,
 			 Saitama, 350-0496, Japan}
\address{	$^2$Department of Physics, Kanagawa Dental College,
                     Kanagawa, 238-8580, Japan}
\date{\today}

\begin{abstract}
We show that the coupling constant of a quantum-induced composite field
	is scale invariant due to its compositeness condition.
It is first demonstrated in next-to-leading order in $1/N$ in typical models,
	and then we argue that it holds exactly.
\end{abstract}

% 11.10.Hi Renormalization group evolution of parameters
% 11.10.Gh Renormalization
% 11.15.Pg Expansions for large numbers of components (e.g., 1/N sub c expansions)
% 12.60.Rc Composite models

\maketitle

Composite fields induced by quantum fluctuations play important roles in 
	wide areas of physics.
The Cooper pair, or the order parameter 
	in the Landau-Ginzberg theory of superconductivity
	can be taken as such a composite of elementary fields in the system.
The ideas are also successfully used to describe properties 
	of hadron dynamics in the Nambu-Jona-Lasinio model (NJL) \cite{NJL}.
They are further applied to 
	induced gauge theories \cite{ig}, 
	induced gravity \cite{iG}, 
	composite models of quarks, leptons, 
	gauge bosons and Higgs scalar \cite{comp},
	various collective motions in nuclei and solid states, 	and 
	brane-induced gravity and field theories \cite{BW} etc. 
The theories of the quantum induced composites 
	are not renormalizable in many cases.
They can, however, be formulated as the special case 
	of some renormalizable theory
	with the compositeness condition (CC) \cite{CC},
	which says that $Z=0$, 
	where $Z$ is the renormalization constant 
	of the to-be-composite field.
For example, with CC, the Yukawa model for elementary fermions and bosons 
	reduces to the NJL model 
	with elementary fermions and composite bosons.

In spite of extensive studies in their long history,
	it is not clear what happens in the infinite cutoff limit,
	or equivalently, integral dimension limit 
	of these non-renormalizable theories \cite{CC,history}.
In fact, in any of known pertuabative treatments, 
	the induced composite coupling constants vanish in this limit.  
Here we do not adhere on this difficult problem.
Instead	we consider non-limiting cases, 
	by fixing the number of spacetime dimensions $d=d_0-2\epsilon$
	at some value close to but different from the integral number of
	physical dimensions $d_0$ (=4 or 6 below).
We interpret that it simulates existing finite cutoffs 
	in various physical systems in the nature.
We adopt the minimal subtraction scheme,
	where the poles in $\epsilon$ are retained 
	in the renormalization constants.
Note that, in this scheme, the renormalization group 
	is well defined even for the non-limiting case $\epsilon\not=0$.

In this paper, we show the following fact: 
{\it In a composite field theory which is equivalent to 
	some renormalizable field theory under CC,
	the induced coupling constant of the composite field
	is exactly scale invariant by virtue of the CC itself}.
We previously demonstrated it in some models at the leading order in $1/N$ 
	where $N$ is the number of the elementary matter species \cite{beta}.
%%% replaced in the revised version %%%
Here we first show it to next-to-leading order in $1/N$
	and to leading order in $\epsilon$ 
	with $g^2\sim\epsilon/N$
	in three typical models. 
%%% original version %%%
%Here we first show it at the leading order in $\epsilon$
%	at the next to leading order in $1/N$ 
%	with $ \epsilon \ll 1/N \ll 1$
%	in three typical models. 
Then we argue that the scale invariance holds exactly \cite{allorder}.
It is remarkable that the awkward non-renormalizable theories
	bury such a high symmetry in their depth!
The compositeness 
	i.\ e.\ absence of its elementary degree of freedom 
	protects the coupling constants
	from flowing with the scale parameter.

\noindent 
A) {\it Scalar composite in six dimensions}\ \ \ 
We consider a system of $2N$ complex scalar fields 
	$\chi_0^i=(\chi_{01}^i,\cdots,\chi_{0N}^i)$ $(i=1,2)$ 
	with mass $m_0$ in six dimensions:
\begin{eqnarray} 
	{\cal L}
	=\Sigma_i(|\partial_\mu\chi_0^i|^2-m_0^2|\chi_0^i|^2)
	-F|(\chi_0^1\chi_0^2)|^2,
	\label{6dL}
\end{eqnarray} 
where $F$ is a coupling constant.
This is one of the simplest models that realize our present idea.
Naively we can see that the chain of  
	$\chi_0^i$-loop diagrams
	gives rise to their composite pole 
	in the total momentum square of the channel.
It is systematically described in the following way.
The Lagrangian (\ref{6dL}) is equivalent to
\begin{eqnarray} 
	{\cal L'}
	&=&\Sigma_i(|\partial_\mu\chi_0^i|^2-m_0^2|\chi_0^i|^2)
\cr&& 
	-|\Phi|^2/F
	+\Phi(\chi_0^1\chi_0^2)+{\rm h.c.},
	\label{6dLaux}
\end{eqnarray} 
where $\Phi$ is an auxiliary field.
We compare this with the renormalizable model 
	for $\chi_0^i$
	and an elementary complex scalar $\phi_0$
	with mass $M_0$ and a coupling constant $g_0$:
\begin{eqnarray} 
	\tilde{\cal L}
	&=&\Sigma_i(|\partial_\mu\chi_0^i|^2-m_0^2|\chi_0^i|^2)
	+|\partial_\mu\phi_0|^2
\cr&&  
	-M_0^2|\phi_0|^2
	+g_0 \phi_0(\chi_0^1\tilde\chi_0^2)+{\rm h.c.}.
	\label{6dLtr}
\end{eqnarray} 
We renormalize $\chi_0^i$, $\phi_0$, $m_0$, $M_0$, and $g_0$ 
	with the renormalization constants 
	$Z_1$, $Z_2$, $Z_3$, $Z_m$, and $Z_M$
	and the renormalized quantities $\chi$, $\phi$, $m$,  $M$ and $g$: 
	$\chi_0^i=\sqrt{Z_2}\chi^i$, 
	$\phi_0=\sqrt{Z_3}\phi$, 
	$\sqrt{Z_2}m_0=\sqrt{Z_m} m$, 
	$\sqrt{Z_3}M_0=\sqrt{Z_M} M$ and
\begin{eqnarray} 
	Z_2\sqrt{Z_3}g_0=Z_1 g\mu^\epsilon, \ 
	\label{6dZ}
\end{eqnarray} 
	where a mass parameter $\mu$ is 
	introduced to make the coupling constant $g$ dimensionless.
We can directly see that (\ref{6dLtr}) entirely coincides with (\ref{6dLaux}) if
\begin{eqnarray} 
	Z_3=0,\ 
	Z_1\not=0,\ 
	Z_2\not=0,\ 
	Z_m\not=0,\ 
	Z_M\not=0,\ 
	\label{6dCC}
\end{eqnarray} 
and if we identify $Z_2\Phi$ with $Z_1g \mu^\epsilon\phi$ 
and $F$ with $Z_1^2g^2\mu^{2\epsilon}/Z_2^2Z_MM^2$.
Thus we can calculate the physical quantities in the system with (\ref{6dL})
	at non-vanishing $\epsilon$ via well understood Lagrangian (\ref{6dLtr})
	with the condition (\ref{6dCC}),
	which is called ``compositeness condition (CC)".
Eq. (\ref{6dZ}) indicates that the CC (\ref{6dCC}) is equivalent to 
\begin{eqnarray} 
	g_0\rightarrow\infty,\ 
	Z_1\not=0,\ 
	Z_2\not=0,\ 
	Z_m\not=0,\ 
	Z_M\not=0.
	\label{6dg0inf}
\end{eqnarray}

After calculations, we get (with $I=1/6(4\pi)^3\epsilon$)
\begin{eqnarray} 
	Z_1&=&1,\ \ \ 
	Z_2=1+{N}^{-1}\ln\left(1-{Ng^2I}\right),
	\label{6dZ12}
\\
	Z_3&=&1-(N+2)g^2I
	\nonumber
\\&&
	-{2N^{-1}}(1-Ng^2I)\ln(1-Ng^2I)
	\label{6dZ3}
\end{eqnarray} 
%%% replaced in the revised version %%%
	to next-to-leading order in $1/N$
	and leading order in $\epsilon$ with $g^2\sim\epsilon/N$. 
%%% original %%%
%	at the leading order in $\epsilon$ 
%	at the next to leading order in $1/N$.
Then, the renormalization group beta function is calculated as
\begin{eqnarray} 
	\beta\equiv\mu{\partial g/\partial \mu}
	=-\epsilon g+(N+2)g^3/6(4\pi)^3.
	\label{6dbeta}
\end{eqnarray} 
The differential equation (\ref{6dbeta}) is solved   
	to give the running coupling constant 
\begin{eqnarray}&& 
	g^2=\left[(N+2)/6(4\pi)^3\epsilon+\mu^{2\epsilon}/g_0^2\right]^{-1}
\cr&&
	=\left[a-(N+2)\ln\mu^2/6(4\pi)^3\right]^{-1}+O(\epsilon), 
	\label{6dg2de}
\end{eqnarray} 
where the integration constant is chosen in accordance with (\ref{6dZ}), and
	$a=\lim_{\epsilon\rightarrow0}\{(N+2)I+1/g_0^2)\}$.

With CC (\ref{6dg0inf}), 
	the $\mu$-dependent part in (\ref{6dg2de}) disappears,
	and (\ref{6dg2de}) reduces to the scale invariant form
\begin{eqnarray} 
	g^2=6(4\pi)^3\epsilon/(N+2).
	\label{6dg2}
\end{eqnarray} 
In fact, (\ref{6dg2}) is the solution 
	of CC (\ref{6dCC}) with $Z_3$ in (\ref{6dZ3}),
	and it implies $\beta=0$
	within the present approximation.
Thus, here, the compositeness implies the scale invariance of 
	the induced coupling constant.

\noindent
B) {\it Nambu-Jona-Lasinio model in four dimensions}\ \ \ 
We consider a system of $N$ fermions
	$\psi_0=(\psi_{01},\cdots,\psi_{0N})$ 
	in four dimensions with the Lagrangian
\begin{eqnarray} 
	{\cal L}
	=\overline\psi_0i\!\not\!\partial\psi_0
	-F|\overline\psi_{0\rm L}\psi_{0\rm R}|^2,
	\label{NJLL}
\end{eqnarray} 
where $F$ is a coupling constant.
(Note that the notations are renewed model by model.)
The composite pole due to the chain of $\psi_0$-loops
	is systematically described in the following way.
The Lagrangian (\ref{NJLL}) is equivalent to
\begin{eqnarray} 
	{\cal L'}
	=\overline\psi_0i\!\not\!\partial\psi_0
	+\overline\psi_{0\rm L}\Phi\psi_{0\rm R}+{\rm h.c.}
	-F^{-1}|\Phi|^2
	\label{NJLLaux}
\end{eqnarray} 
where $\Phi$ is an auxiliary field.
We compare this with the Yukawa model 
	for $\psi_0$ and an elementary boson $\phi_0$
	with mass $M_0$ and coupling constants $g_0$ and $\lambda _0$:
\begin{eqnarray} 
	\tilde{\cal L}&= \overline \psi_0 i\!\not\!\partial \psi_0
	+ g_0(\overline \psi _{0\rm L}\phi _0\psi _{0\rm R}+{\rm h.c.})
\nonumber
\\&
	+ |\partial _\mu \phi_0|^2
	- M_0^2|\phi_0|^2
	- \lambda_0|\phi_0|^4	
	\label{NJLLtr}
\end{eqnarray} 
We renormalize $\psi_0$, $\phi_0$, $M_0$, $g_0$ and $\lambda_0$
	with the renormalization constants $Z_1$, $Z_2$, $Z_3$, $Z_4$, and $Z_M$ 
	and the renormalized quantities 	$\psi$, $\phi$, $M$, $g$ and $\lambda$:  
	$\psi_0=\sqrt{Z_2}\psi$, 
	$\phi_0=\sqrt{Z_3}\phi$, 
	$\sqrt{Z_3}M_0=\sqrt{Z_M} M$ and
\begin{eqnarray} 
	Z_2\sqrt{Z_3}g_0=Z_1 g\mu^\epsilon, \ 
	Z_3^2\lambda_0=Z_4 \lambda\mu^{2\epsilon}, \ 
	\label{NJLZ}
\end{eqnarray} 
	where the coupling constants $g$ and $\lambda$ are made dimensionless
	with a mass parameter $\mu$.
We can see that (\ref{NJLLtr}) entirely coincides with (\ref{NJLLaux}) 
	if we have the CC
\begin{eqnarray} 
	Z_3=0,\  
	Z_4=0,\ 
	Z_1\not=0,\ 
	Z_2\not=0,\ 
	Z_M\not=0,\ 
	\label{NJLCC}
\end{eqnarray} 
and we identify $F$ with $Z_1^2g^2\mu^{2\epsilon}/Z_2^2Z_MM^2$ and 
	$Z_2\Phi$ with $Z_1g\mu^\epsilon\phi$.
Thus we can calculate the physical quantities in the system with (\ref{NJLL})
	at non-vanishing $\epsilon$ via well understood Lagrangian (\ref{NJLLtr})
	with CC (\ref{NJLCC}).
Eq. (\ref{NJLZ}) indicatse that the CC (\ref{NJLCC}) is equivalent to
\begin{eqnarray}
	g_0\rightarrow\infty,\ 
	Z_1\not=0,\ 
	Z_2\not=0,\ 
	Z_M\not=0 
	\label{NJLg0inf}
\end{eqnarray}
with arbitrary $\lambda_0$

After calculations, we get (with $I=1/(4\pi)^2\epsilon$)
\begin{eqnarray}&&\hspace{-15pt}
	Z_1=1,\ \ \ 
	Z_2=1+\ln(1-Ng^2I),
	\label{NJLZ12}
\\&&\hspace{-15pt}
	Z_3=1-(N+1)g^2I
	-N^{-1} (1-Ng^2I)\ln(1-Ng^2I)
	\nonumber
\\&&\hspace{-15pt}
	Z_4=1-(N-8)g^2I/\lambda
	-20(\lambda-g^2)^2I/\lambda(1-Ng^2I)
	\nonumber
\\&&\hspace{-15pt}
	-(N\lambda)^{-1}\left(20\lambda-{18g^2}-{2Ng^4I}\right)
	\ln\left(1-{Ng^2I}\right)
	\label{NJLZ34}
\end{eqnarray} 
%%% replaced in the revised version %%%
	to next-to-leading order in $1/N$
	and leading order in $\epsilon$ with $g^2\sim\epsilon/N$. 
%%% original %%%
%	at the leading order in $\epsilon$ 
%	at the next to leading order in $1/N$.
Then the renormalization group beta functions are calculated as 
\begin{eqnarray} &&\hspace{-15pt}
	\beta_g\equiv\mu\partial g/\partial \mu
	=-\epsilon g+(N+1)g^3/(4\pi)^2.
\cr&&\hspace{-15pt}
	\beta_\lambda\equiv\mu\partial \lambda/\partial \mu
	=-2\epsilon \lambda
	+(4N\lambda g^2
	-2Ng^4)/(4\pi)^2
\cr&&\ \ \ \ \ \ \ \ \ \ \ \ \ \ \ 
	+(40\lambda^2
	-40\lambda g^2
	+20g^4)/(4\pi)^2
	\label{NJLbeta}
\end{eqnarray} 
The coupled differential equation (\ref{NJLbeta}) is solved 
	to give the running coupling constants
\begin{eqnarray} &&\hspace{-15pt}
	g^2=1/[(N+1)I+\mu^{2\epsilon}/g_0^2],
\cr&&\hspace{-15pt}
	\lambda=\frac{(N-8)I+\lambda_0\mu^{2\epsilon}/g_0^4}
	{((N+1)I+\mu^{2\epsilon}/g_0^2)^2}
	-\frac{20(\lambda_0-g_0^2)^2I\mu^{2\epsilon}}	
	{(Ng_0^2I+\mu^{2\epsilon})^3}
\cr&&  
	-\frac{18(\lambda_0-g_0^2)\mu^{2\epsilon}}	
	{N(Ng_0^2I+\mu^{2\epsilon})^2}
	\ln{\left(1+{Ng_0^2\mu^{-2\epsilon}I}\right)}
	\label{NJLg2de}
\end{eqnarray} 
where the integration constants are chosen with (\ref{NJLZ}).

With CC (\ref{NJLg0inf}), 
	the $\mu$-dependent parts in (\ref{NJLg2de}) disappear,
	and (\ref{NJLg2de}) reduces to the scale invariant form 
\begin{eqnarray} 
	g^2=(4\pi)^2\epsilon/(N+1),\ \ \ \ 
	\lambda=(4\pi)^2\epsilon/(N+10).
	\label{NJLg2}
\end{eqnarray} 
In fact, (\ref{NJLg2}) is the solution 
	of CC (\ref{NJLCC}) with $Z_3$ and $Z_4$ in (\ref{NJLZ34}) \cite{NJLNL},
	and it implies $\beta_g=0$ and  $\beta_\lambda=0$
	within the present approximation.
Thus, here, the compositeness implies the scale invariance of 
	the induced coupling constants.

\noindent
C) {\it Induced gauge theory in four dimensions}\ \ \ 
We consider the strong coupling limit $F\rightarrow\infty$
	of the system of $N$ $SU(N_c)$-$N_c$-plet fermions 
	$\psi_0=(\psi_{01},$ $\cdots,\psi_{0N})$ with mass $m_0$
 	in four dimensions:
\begin{eqnarray} 
	{\cal L}
	=\overline\psi_0(i\!\not\!\partial-m_0)\psi_0
	-F(\overline\psi_{0}T^a\gamma_\mu\psi_{0})^2,
	\label{indL}
\end{eqnarray} 
where $T^a$ is the generator matrix 
	of the fundamental representation of $SU(N_c)$, and
	$\gamma_\mu$ is the Dirac matrix.
(Note that the notations are renewed model by model.)
The composite pole due to the chain of $\psi_0$-loops 
	is systematically described in the following way.
The Lagrangian (\ref{indL}) is equivalent to
\begin{eqnarray} 
	{\cal L'}
	=\overline\psi_0(i\!\not\!\partial-m_0)\psi_0
	+\overline\psi_0T^a\!\not\!\Phi^a\psi_0
	-F^{-1}|\Phi_\mu^a|^2
	\label{indLaux}
\end{eqnarray} 
where $\Phi_\mu^a$ is an auxiliary vector field 
	in the adjoint representation of $SU(N_c)$.
We compare this with the $SU(N_c)$ gauge theory 
	for $\psi_0$ and 
	the elementary gauge boson $G_{0\mu}^a$
	with the gauge coupling constant $g_0$
\begin{eqnarray} &&\hspace{-15pt}
	\tilde{\cal L}=
	\overline\psi_0 (i\!\not\!\!\partial-m_0)\psi_0
 	+ g_0\overline\psi_0 T^a\!\not\!\!G_0^a\psi_0
	-\frac{1}{4}(G_{0\mu\nu}^a)^2,
	\label{indLtr}
\end{eqnarray} 
where $G_{0\mu\nu}^a$ is the field strength of $G_{0\mu}^a$. 
We renormalize $\psi_0$, $G_{0\mu}^a$, $M_0$ and $g_0$ 
	with the renormalization constants $Z_1$, $Z_2$, $Z_3$ and $Z_m$  
	and the renormalized quantities $\psi$, $G_\mu^a$, $M$ and $g$:
	$\psi_0=\sqrt{Z_2}\psi$, 
	$G_{0\mu}^a=\sqrt{Z_3}G_\mu^a$, 
	$Z_2m_0=Z_m m$ and
\begin{eqnarray} 
	Z_2\sqrt{Z_3}g_0=Z_1 g\mu^\epsilon, \ 
	\label{indZ}
\end{eqnarray} 
	where the coupling constant $g$ is made dimensionless
	with a mass parameter $\mu$.
We can see that (\ref{indLtr}) entirely coincides with (\ref{indLaux}) 
	if we have the CC
\begin{eqnarray}  
	Z_3=0,\ \ \ 
	Z_1\not=0,\ \ \ 
	Z_2\not=0,\ \ \ 
	Z_m\not=0,\ \ \ 
	\label{indCC}
\end{eqnarray} 
	and we take $Z_2\Phi_\mu^a=Z_1g\mu^\epsilon G_\mu^a$ 
	and $F\rightarrow\infty$.
Thus we can calculate the physical quantities in the system with (\ref{indL})
	at non-vanishing $\epsilon$ 
	via well understood Lagrangian (\ref{indLtr})
	with CC (\ref{indCC}).
Note that, for a consistent quantum description,
	we should and we can introduce the gauge fixing term 
	and the Fadeev-Popov term
	without changing physical contents.
We denote the gauge parameter by $\alpha$.
Eq. (\ref{indZ}) indicates that the CC is equivalent to 
\begin{eqnarray} 
	g_0\rightarrow\infty,\ \ \ 
	Z_1\not=0,\ \ \ 
	Z_2\not=0,\ \ \ 
	Z_m\not=0,\ \ \ 
	\label{indg0inf}
\end{eqnarray} 
in terms of the bare parameters of (\ref{indLtr}).

After calculations, we obtain (with $I=1/(4\pi)^2\epsilon$)
\begin{eqnarray}&& \hspace{-15pt}
	Z_1=1+\frac{9N_c}{8N}\ln\left(1-{2Ng^2I\over3}\right)
	-{\alpha g^2I}\frac{3N_c^2-2}{4N_c}, 
\nonumber\\&&\hspace{-15pt}
	Z_2=1-{\alpha g^2I}{(N_c^2-1)/2N_c},  \ \ \  \ \ 
	\label{indZ12}
\\&&\hspace{-15pt}
	Z_3=1
	-\frac{(2N-11N_c) g^2I}{3}
	-\frac{\alpha N_c g^2I}{2}
	\left(1-\frac{2N g^2I}{3}\right)  
\cr&& 
	+\frac{9N_c}{4N}\left(1-\frac{2Ng^2I}{3}\right)
	\ln\left(1-\frac{2Ng^2I}{3}\right)
	\label{indZ3}
\end{eqnarray} 
%%% replaced in the revised version %%%
	to next-to-leading order in $1/N$
	and leading order in $\epsilon$ with $g^2\sim\epsilon/N$. 
%%% original %%%
%	at the leading order in $\epsilon$ 
%	at the next to leading order in $1/N$.
Then, the renormalization group beta function is calculated as
\begin{eqnarray} 
	\beta \equiv\mu{\partial g/\partial \mu}
	=-\epsilon g+{(2N-11N_c)g^3/3(4\pi)^2}.
	\label{indbeta}
\end{eqnarray} 
The differential equation (\ref{indbeta}) is solved   
	to give the running coupling constant 
\begin{eqnarray} &&\hspace{-15pt}
	g^2=[{(2N-11N_c)/3(4\pi)^2\epsilon+\mu^{2\epsilon}/g_0^2}]^{-1}
\nonumber\\&&\hspace{-10pt}
	\left(=[a-(2N-11N_c)\ln\mu^2/3(4\pi)^2]^{-1}+O(\epsilon)\right)
	\label{indg2de}
\end{eqnarray} 
where the integration constant is chosen with (\ref{indZ}), and
	$a=\lim_{\epsilon\rightarrow0}\{(2N-11N_c)I/3 +1/g_0^2)\}$.

With CC (\ref{indg0inf}), 
	the $\mu$-dependent part in (\ref{indg2de}) disappears,
	and (\ref{indg2de}) reduces to the scale invariant form 
\begin{eqnarray} 
	g^2={3(4\pi)^2\epsilon}/({2N-11N_c}).\ \ \ \ 
	\label{indg2}
\end{eqnarray} 
In fact, (\ref{indg2}) is the solution 
	of CC (\ref{indCC}) with $Z_3$ in (\ref{indZ3}),
	and it implies $\beta =0$
	within the present approximation.
Thus, here again, the compositeness implies the scale invariance of 
	the induced coupling constant.

We argue that the scale invariance of the composite-field
	coupling constants persists in all orders.
The $\mu$ dependences of $g$'s and $\lambda$ 
	in (\ref{6dg2de}), (\ref{NJLg2de}) and (\ref{indg2de})
	due to the differential equations 
	(\ref{6dbeta}), (\ref{NJLbeta}) and (\ref{indbeta})
	originally come from the relations
	(\ref{6dZ}), (\ref{NJLZ}) and (\ref{indZ}).
The solutions of (\ref{6dbeta}), (\ref{NJLbeta}) and (\ref{indbeta})
	are given by the algebraic solutions of the equations
	(\ref{6dZ}), (\ref{NJLZ}) and (\ref{indZ}) with 
	$Z$'s in
	(\ref{6dZ12}), (\ref{6dZ3}), (\ref{NJLZ12}), (\ref{NJLZ34}), 
	(\ref{indZ12}) and (\ref{indZ3}) inserted. 
%The renormalization constants do not explicitly depend on $\mu$
%	(by definition in the minimal subtraction scheme,
%	and, in the momentum subtraction scheme, 
%	up to $O(m^2/\mu^2)$ which we neglect). 
The $\mu$-dependences of the coupling constants
	as the algebraic solutions of 
	equations (\ref{6dZ}), (\ref{NJLZ}) and (\ref{indZ})
	arise through the factors $\mu^\epsilon$ and  $\mu^{2\epsilon}$
	in the right-hand side of (\ref{6dZ}), (\ref{NJLZ}) and (\ref{indZ}).
With CC (\ref{6dg0inf}), (\ref{NJLg0inf}) and (\ref{indg0inf}),
	the $\mu$-dependent parts disappear from
	the equations (\ref{6dZ}), (\ref{NJLZ}) and (\ref{indZ})
	in all orders.
Therefore the coupling constants $g$'s and $\lambda$ with CC
	are independent of the scale $\mu$ in all orders.
The scale invariance of the composite-field coupling constants holds exactly.

It holds in any order as far as the expansion is proper
	as in the case of the $1/N$-expansion.
However, the coupling-constant expansions and loop expansions 
	are improper under CC, 
	because CC at the lowest two orders implies 
	that infinite higher diagrams have the same order of magnitude,
	and we cannot see the CC based scale invariance 
	at any particular order in these expansions.
In general, the CC can have isolated solution that is not a limit 
	with respect to some particular perturbative parameter.
The CC based scale invariance persists even in this non-perturbative case. 
The composite coupling constants are entirely on the fixed point
	from infrared to ultraviolet region.
They neither are asymptotically free nor blow up at some finite scale.
The composite field coupling constant has no Landau pole.
%All of these properties are proved for non-vanishing fixed $\epsilon$
%	i.\ e.\ for finite cutoff $\Lambda\sim me^{-1/\epsilon}$, 
%	but not for infinite cutoff.
%Therefore all of them are realized only for $m\ll\mu\ll\Lambda$.
So far we considered the cases where 
	the induced coupling constants are dimensionless. 
It is straightforward to extend the present argument
 	to the cases of induced coupling constants 
	with positive mass dimensions, i.e. 
	the composite theories equivalent to super-renormalizable   
	theories with CC.
The induced composite-field coupling constants 
	would scale with its canonical dimensions.

The scaling properties would provide a powerful clue 
	to discriminate between compositeness and elementariness 
	phenomenologically.
For example, if the weak bosons and Higgs scalar 
	are quantum-induced composites \cite{comp}, 
	the coupling constants should be scale invariant. 
On the other hand, 
	the photon and gluons cannot be quantum-induced composites
	in spite of several theoretical suggestions \cite{ig,compAG},
	since running of coupling constants are supported by experiments.
The elementariness of the gluon is consistent with
	the related theoretical indication by complementarity 
	that asymptotically free gauge bosons 
	cannot be a quantum-induced composite \cite{complementarity}.
%%%%% replaced in revised version %%%%%
The scaling behavior would also be realized 
	in various phenomena based on collective motions 
	in condensed matters.
When fundamental fermions are coupled to gravity, 
	the dimensionless part of the induced composite scalar sector 
%1	the induced composite scalar sector 
	is known to exhibit conformal symmetry, 
	reflecting asymptotic conformal invariance 
%1	owing to asymptotic conformal invariance 
	of the fermion sector in the ultraviolet region 
	\cite{conformal}.
It would be an interesting challenge to inquire the interrelations between
	this conformal symmetry and the scale invariance demonstrated here.
%1	this conformal symmetry and the scale invariance presented here.
The idea of the quantum-induced composite is used explicitly or implicitly
	in many areas from particle to cosmological physics.
We expect that the scaling properties presented here
%1 We expect that the scaling properties demonstrated here
	would have chances to be realized in many of such systems
	and would elucidate varieties of phenomena in the nature.
%%%%% original version %%%%%%%%%%
%We suppose that the scaling behavior would be realized 
%	in various phenomena based on collective motions 
%	in nuclei and solid states
%	because they are taken as kinds of induced composite fields.
%The idea of the quantum-induced composite is used explicitly or implicitly
%	in many areas from particle to cosmological physics.
%We expect that the scaling properties presented here
%	would have chances to be realized in many of such systems
%	and would elucidate varieties of phenomena in the nature.

% cutoff
% Landau pole, no blow up, IR or UV fixed point
% useful weak boson
% application to solid states
%

% g expansion, loop expansion
% renormalizability problem
% exact equivalence 

% Landau pole, no blow up, IR or UV fixed point
% difference from Eguchi 
% 
% new type of scaling protection mechanism

% beta=0 ---> Z=0

% infinite cutoff
% general proof
% useful weak boson
% application to solid states

\end{document}